\documentclass[12pt]{article}
\usepackage{html}
\usepackage{epsf}
\usepackage[pdftex]{graphicx}
\usepackage{graphicx}
\usepackage{amssymb}
\usepackage{amsmath}
\usepackage{hyperref}
\begin{document}
\begin{titlepage}
\includegraphics[width=150mm]{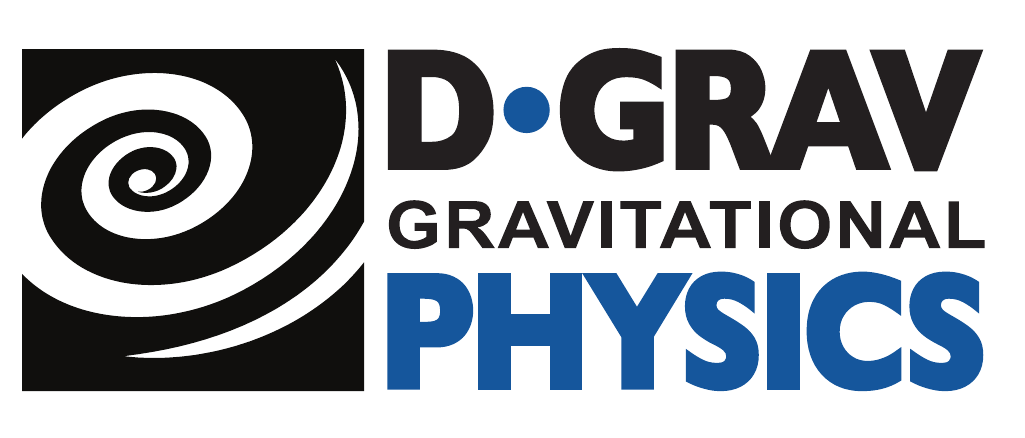}
\begin{center}
{ \Large {\bf MATTERS OF GRAVITY}}\\ 
\bigskip
\hrule
\medskip
{The newsletter of the Division of Gravitational Physics of the American Physical 
Society}\\
\medskip
{\bf Number 53 \hfill June 2019}
\end{center}
\begin{flushleft}
\tableofcontents
\end{flushleft}
\end{titlepage}
\vfill\eject
\begin{flushleft}
\section*{\noindent  Editor\hfill}
David Garfinkle\\
\smallskip
Department of Physics
Oakland University
Rochester, MI 48309\\
Phone: (248) 370-3411\\
Internet: 
\htmladdnormallink{\protect {\tt{garfinkl-at-oakland.edu}}}
{mailto:garfinkl@oakland.edu}\\
WWW: \htmladdnormallink
{\protect {\tt{http://www.oakland.edu/physics/Faculty/david-garfinkle}}}
{http://www.oakland.edu/physics/Faculty/david-garfinkle}\\

\section*{\noindent  Associate Editor\hfill}
Greg Comer\\
\smallskip
Department of Physics and Center for Fluids at All Scales,\\
St. Louis University,
St. Louis, MO 63103\\
Phone: (314) 977-8432\\
Internet:
\htmladdnormallink{\protect {\tt{comergl-at-slu.edu}}}
{mailto:comergl@slu.edu}\\
WWW: \htmladdnormallink{\protect {\tt{http://www.slu.edu/arts-and-sciences/physics/faculty/comer-greg.php}}}
{http://www.slu.edu/arts-and-sciences/physics/faculty/comer-greg.php}\\
\bigskip
\hfill ISSN: 1527-3431


\bigskip

DISCLAIMER: The opinions expressed in the articles of this newsletter represent
the views of the authors and are not necessarily the views of APS.
The articles in this newsletter are not peer reviewed.

\begin{rawhtml}
<P>
<BR><HR><P>
\end{rawhtml}
\end{flushleft}
\pagebreak
\section*{Editorial}

The next newsletter is due December 2019.  Issues {\bf 28-53} are available on the web at
\htmladdnormallink 
{\protect {\tt {https://files.oakland.edu/users/garfinkl/web/mog/}}}
{https://files.oakland.edu/users/garfinkl/web/mog/} 
All issues before number {\bf 28} are available at
\htmladdnormallink {\protect {\tt {http://www.phys.lsu.edu/mog}}}
{http://www.phys.lsu.edu/mog}

Any ideas for topics
that should be covered by the newsletter should be emailed to me, or 
Greg Comer, or
the relevant correspondent.  Any comments/questions/complaints
about the newsletter should be emailed to me.

A hardcopy of the newsletter is distributed free of charge to the
members of the APS Division of Gravitational Physics upon request (the
default distribution form is via the web) to the secretary of the
Division.  It is considered a lack of etiquette to ask me to mail
you hard copies of the newsletter unless you have exhausted all your
resources to get your copy otherwise.

\hfill David Garfinkle 

\bigbreak

\vspace{-0.8cm}
\parskip=0pt
\section*{Correspondents of Matters of Gravity}
\begin{itemize}
\setlength{\itemsep}{-5pt}
\setlength{\parsep}{0pt}
\item Daniel Holz: Relativistic Astrophysics,
\item Bei-Lok Hu: Quantum Cosmology and Related Topics
\item Veronika Hubeny: String Theory
\item Pedro Marronetti: News from NSF
\item Luis Lehner: Numerical Relativity
\item Jim Isenberg: Mathematical Relativity
\item Katherine Freese: Cosmology
\item Lee Smolin: Quantum Gravity
\item Cliff Will: Confrontation of Theory with Experiment
\item Peter Bender: Space Experiments
\item Jens Gundlach: Laboratory Experiments
\item Warren Johnson: Resonant Mass Gravitational Wave Detectors
\item David Shoemaker: LIGO 
\item Stan Whitcomb: Gravitational Wave detection
\item Peter Saulson and Jorge Pullin: former editors, correspondents at large.
\end{itemize}
\section*{Division of Gravitational Physics (DGRAV) Authorities}
Chair: Gary Horowitz; Chair-Elect: Nicolas Yunes
; Vice-Chair: Gabriela Gonzalez . 
Secretary-Treasurer: Geoffrey Lovelace; Past Chair: Emanuele Berti ; Councilor: Beverly Berger
Members-at-large:
Lisa Barsotti, Theodore Jacobson, Michael Lam, Jess McIver, Alessandra Corsi, Henriette Elvang.
Student Members: Belinda Cheeseboro, Alejandro Cardenas-Avendano.
\parskip=10pt

\vfill\eject

\section*{\centerline
{we hear that \dots}}
\addtocontents{toc}{\protect\medskip}
\addtocontents{toc}{\bf DGRAV News:}
\addcontentsline{toc}{subsubsection}{
\it we hear that \dots , by David Garfinkle}
\parskip=3pt
\begin{center}
David Garfinkle, Oakland University
\htmladdnormallink{garfinkl-at-oakland.edu}
{mailto:garfinkl@oakland.edu}
\end{center}

Clifford Will has been awarded the 2019 Albert Einstein Medal by the Albert Einstein Society in Berne, Switzerland.

Gabriela Gonzalez has been named the 2019 SEC Professor of the Year.

Bernard Schutz has been elected to the National Academy of Sciences.

Gabriela Gonzalez has been elected Vice-Chair of DGRAV.  Alessandra Corsi and Henriette Elvang have been elected Members-at-large of the DGRAV Executive Committee.  Alejandro Cardenas-Avendano has been elected Student Student Member of the DGRAV Executive Committee.

Hearty Congratulations!

\vfill\eject

\section*{\centerline
{Gravitational-wave Standard Sirens}}
\addtocontents{toc}{\protect\medskip}
\addtocontents{toc}{\bf Research Briefs:}
\addcontentsline{toc}{subsubsection}{
\it  Gravitational-wave Standard Sirens, by Daniel Holz and Maya Fishbach}
\parskip=3pt
\begin{center}
Daniel Holz, University of Chicago
\htmladdnormallink{holz-at-uchicago.edu}
{mailto:holz@uchicago.edu}\\
Maya Fishbach, University of Chicago
\htmladdnormallink{mfishbach-at-uchicago.edu}
{mailto:mfishbach@uchicago.edu}
\end{center}

\subsection*{Overview}


Over thirty years ago Bernie Schutz pointed out that general relativity provides an exceedingly elegant way to determine the absolute distance to a source at cosmological distances~\cite{schutz}. By measuring the gravitational waves from the inspiral and merger of two compact objects, such as neutron stars or black holes, one can infer both the mass scale and the luminosity distance to the source. The scale, known as the chirp mass, comes from the frequency evolution,  while the distance comes from the measured amplitude of the waveform in the detectors~[see]\cite{2018PhT....71l..34H}[for an introductory discussion]. These sources are the gravitational analog of ``standard candles'' such as Type Ia supernovae, which can be used to infer absolute distances through use of a cosmological version of the inverse square law. Because of this close correspondence, these binary sources have been dubbed ``standard sirens''~\cite{2005ApJ...629...15H}.
Unlike in the case of standard candles, 
standard sirens offer a  direct and independent way to measure distances, completely obviating the need for a distance ladder. They are calibrated directly by general relativity.

An absolute distance measurement is of particular interest when coupled with a redshift measurement to the same source. This combination constrains the distance-redshift relation, which is a measure of the scale of the universe as a function of time. Measurements of this relation help determine interesting quantities such as the age and composition of the universe and the nature of the dark energy. At low redshift, this relation can be approximated by $cz=H_0d$, with $c$ the speed of light, $z$  the redshift, $H_0$  the Hubble constant, and $d$  the distance to the source. A standard siren provides $d$ directly from the amplitude of the gravitational waves, and with a measurement of the redshift to the source one can directly infer the Hubble constant.



It is to be emphasized, however, that the redshift of the source cannot be straightforwardly determined from gravitational waves alone. There is a redshift degeneracy: a lower mass binary farther away will be redshifted to have an identical (modulo amplitude) waveform to a closer, more massive binary. The simplest way to address this is to find an electromagnetic counterpart to the gravitational-wave source, and measure the redshift using photons. We call this the ``counterpart standard siren'' approach~\cite{2005ApJ...629...15H}, and short gamma-ray bursts have been thought to be a particularly promising counterpart source~\cite{2006PhRvD..74f3006D,2010ApJ...725..496N,2013arXiv1307.2638N}. An alternate ``statistical standard siren'' approach is to identify  potential host galaxies within the gravitational-wave localization volume, with each galaxy providing a potential redshift to the source~\cite{schutz,macleod,delpozzo}.
Additionally, structure present in the population of sources, such as known features in the mass distribution, can be used in a similar manner~\cite{1996PhRvD..53.2878F,2012PhRvD..85b3535T}. 
Finally, it may be possible to use properties of the source (such as knowledge of the equation-of-state of neutron stars) to extract frequency-dependent features in the waveform and thereby determine redshift directly~\cite{2012PhRvL.108i1101M,2017PhRvD..95d3502D}.

For thirty years after Bernie's original paper, a number of obstinate souls toiled at developing the standard siren approach to cosmology. In the absence of the detection of gravitational waves, much less the detection of a gravitational wave source with an associated electromagnetic counterpart allowing for a redshift determination, the field of standard siren science remained somewhat speculative and quixotic. This all changed in August of 2017.

\subsection*{GW170817}
\begin{figure}
    \centering
    \includegraphics[width=0.5\textwidth]{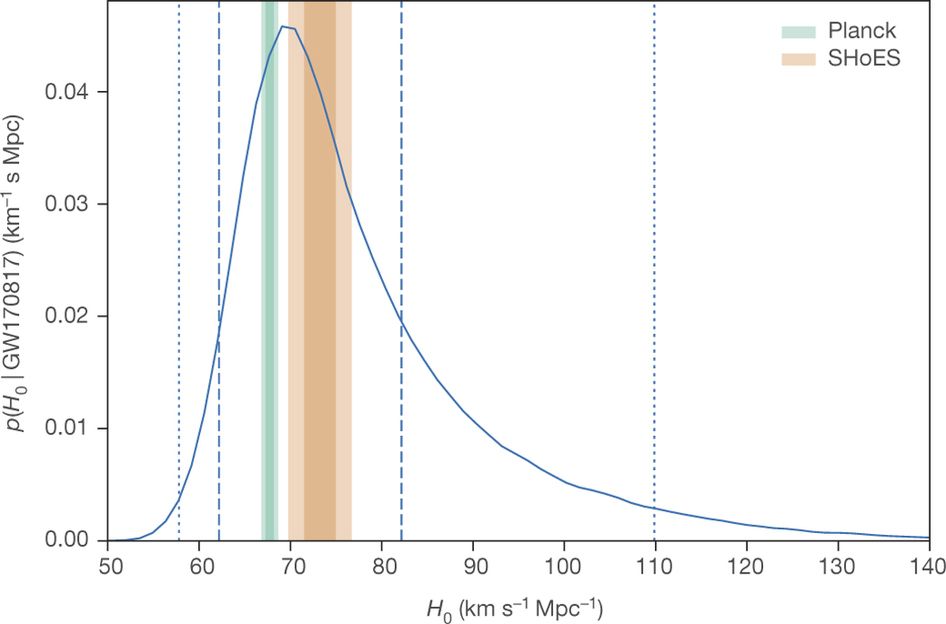}
\caption{Posterior probability density for the Hubble constant given the gravitational-wave distance measurement from GW170817 and the redshift of its host galaxy (solid blue line). The dashed and dotted lines show minimal 68.3\% and 95.4\% intervals. For comparison, the shaded green and orange bands show the 1$\sigma$ (dark shading) and 2$\sigma$ (light shading) constraints from the CMB~\cite{Planck:2015} and from Type Ia supernovae~\cite{shoes}, respectively. Reproduced from~\cite{2017Natur.551...85A}}
    \label{fig:GW170817counterpart}
\end{figure}

GW170817 was detected shortly before 8am Chicago time on August 17, 2017~\cite{PhysRevLett.119.161101}. Less than a day later we were making the first standard siren measurement of the Hubble constant!
This binary neutron-star merger was the first gravitational-wave source with a detected electromagnetic counterpart. Two seconds following the merger, the Fermi gamma-ray space telescope detected a short gamma-ray burst spatially coincident with the gravitational-wave localization error box on the sky~\cite{2017ApJ...848L..13A}. While it was incredibly exciting to detect gamma-rays and gravitational-waves from the same source for the first time, the gamma-ray burst was poorly localized on the sky.
Over the next day, an extensive multi-messenger campaign was launched~\cite{2017ApJ...848L..12A}, and eleven hours later, an optical transient was detected in the galaxy NGC~4993~\cite{Coulter1556,2017ApJ...848L..16S}. Further study revealed that this optical transient was a kilonova powered by r-process decay and unambiguously associated with the neutron star merger.
It was now possible to associate a host galaxy with the gravitational-wave source, and thereby determine the redshift and make the very first standard siren measurement of the Hubble constant~\cite{2017Natur.551...85A}. The resulting Hubble constant measurement is shown in Figure~\ref{fig:GW170817counterpart}. With a single source, the Hubble constant is constrained to roughly $\sim15\%$: $70^{+12}_{-8}$ km/s/Mpc. This first result agrees with the cosmic microwave background measurement from Planck~\cite{Planck:2015} as well as the local distance ladder measurement using Type Ia supernovae from SHoES~\cite{shoes}, which both lie near the peak of the posterior probability shown in Figure~\ref{fig:GW170817counterpart}. As we discuss in the next section, combining this result with future standard sirens will tighten the gravitational-wave measurement and may help shed light on the current Hubble constant tension. We note that the broad distribution in Figure~\ref{fig:GW170817counterpart} is due in part to the distance-inclination degeneracy: a close edge-on binary will be detected with a similar amplitude to a more distance face-on binary~[see, e.g.,]\cite{2006PhRvD..74f3006D,2011CQGra..28l5023S,2017Natur.551...85A}. Independent determinations of inclination would improve the measurement of $H_0$~[see, e.g.,]\cite{2017ApJ...851L..36G,2018arXiv180610596H}, at the (significant) cost of introducing potential astrophysical systematic errors.

The measurements discussed above presume that the theory of general relativity is correct. Alternatively, one can use the standard siren approach to constrain deviations from GR~\cite{Nishizawa2017,Belgacem2017,Amendola2017,Linder2018}. For example, by comparing the measured amplitude of gravitational waves with the inferred luminosity distance through electromagnetic observations, it is possible to constrain the number of spacetime dimensions~\cite{2018JCAP...07..048P} and the running of the Planck mass~\cite{2019PhRvD..99h3504L}.

GW170817 is the only gravitational-wave source with a counterpart thus far, and therefore provides the only existing counterpart standard siren measurement. As a proof of principle, we can pretend that we were unable to identify the kilonova, and thereby unable to determine the host galaxy of GW170817 and determine its redshift. In this case, we would be relegated to the statistical standard siren approach discussed above, where we use every galaxy in the GW170817 localization volume as a potential host galaxy. We followed this approach in~\cite{2019ApJ...871L..13F}, where we combined the gravitational-wave measurements with the GLADE galaxy catalog~\cite{2018MNRAS.479.2374D} to determine the Hubble constant. Since GW170817 is the closest and best-localized gravitational-wave detection, 
the galaxy distribution within the relatively small volume is dominated by a single group of galaxies (of which NGC 4993 is a member), with all of these galaxies at similar redshifts. As a result, the statistical standard siren measurement with GW170817 recovers the same distinct peak at $H_0 \sim 70$ km/s/Mpc which is seen in the counterpart result, with an error that is only roughly twice as broad, $H_0 = 76^{+48}_{-23}$ km/s/Mpc. For comparison, the statistical measurement for an ``average'' detected binary neutron star, at a typical distance and detected with a typical signal-to-noise ratio, would be expected to be $\sim3$ times less informative than the counterpart measurement~\cite{chen17}.


The first application of the statistical standard siren method to a ``dark siren'' was with the binary black hole merger GW170814. This source was the first signal detected by three gravitational-wave detectors, the two LIGO detectors as well as the Virgo detector, and for this reason it was relatively well-localized to $\sim60 \mbox{ deg}^2$ on the sky. The localization region of GW170814 also happened to be in the middle of the footprint of the Dark Energy Survey, which means there happened to be a pre-existing relatively deep and complete catalog of galaxies. GW170814 was therefore an ideal dark standard siren, and we performed a statistical standard siren measurement of the Hubble constant using GW170814 and the DES galaxy catalog in~\cite{2019ApJ...876L...7S}. The resulting measurement is only marginally more informative than the prior: $H_0={78}^{+96}_{−24}$ km/s/Mpc. The weaker constraint from GW170814 compared to Gw170817 is to be expected: because binary black holes are detected at greater distances with larger localization volumes and a significantly greater numbers of potential host galaxies, the Hubble constant measurement from a binary black hole standard siren is typically much less informative than the case of a binary neutron star, even if the neutron stars don't have a counterpart and are analyzed statistically~\cite{chen17}.

\subsection*{The future of standard siren science}

\begin{figure}[t!]
    \centering
    \includegraphics[width=0.5\textwidth]{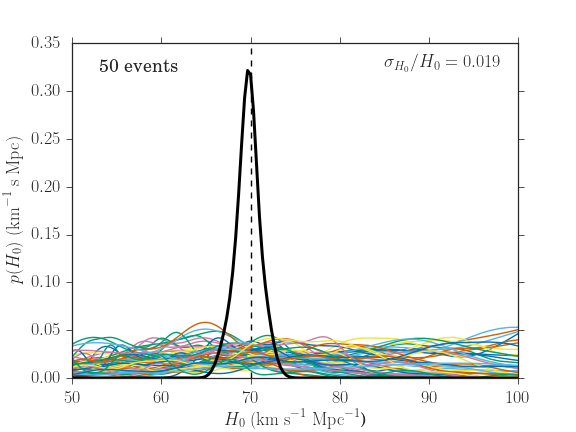}
    \caption{Posterior density on the Hubble constant from 50 simulated binary neutron star events with an associated counterpart (colored lines) together with the combined posterior (black line). The injected value, $H_0 = 70$, km/s/Mpc is shown as the dashed vertical line. With 50 events, we expect to attain  a $\sim2$\% measurement on the Hubble constant.}
    \label{fig:H0combined}
\end{figure}

GW170817 heralds the birth of the field of standard siren cosmology, with this first measurement providing a $\sim15\%$ determination of the Hubble constant. This is an impressive beginning, especially given that it is based on the detection of a single source. However, to make interesting contributions to cosmology, and shed light on the ongoing ``tension'' between early (CMB) and late (supernova) measurements of the Hubble constant, it will be necessary for standard siren measurements to improve to $\lesssim 3\%$~\cite{2018PhRvD..98h3523D}.
This precision is attainable with the gravitational-wave detection of $\sim30$ binary neutron star coalescences~\cite{2013arXiv1307.2638N,chen17,2018arXiv180203404F}. 
Figure~\ref{fig:H0combined} provides an example of a 2\% measurement of the Hubble constant  from 50  binary neutron star detections with counterparts, based on mock data from the ``First 2 Years'' simulated gravitational-wave dataset~\cite{2014ApJ...795..105S}.

The third LIGO/Virgo observational run commenced this past April, and is expected to last $\sim1$ year and produce a few detections of binary neutron stars. The detectors will then be upgraded, and roughly a year later a fourth observational run will commence at the design sensitivity, which is expected to be a $\sim50\%$ improvement in distance over the current configuration~\cite{ObsScen}. After one year of observation with this improved network, neutron-star standard sirens may constrain the Hubble constant to $\sim2\%$~\cite{chen17}; the precise constraint will depend on a number of factors, including the detection rate and the fraction with electromagnetic counterparts and associated redshifts. Roughly speaking, the precision on $H_0$ will scale as $\sim15\%/\sqrt{N}$, where $N$ is the number of detected binary neutron-star sources with counterparts. Without counterparts, the statistical standard siren method yields $\sim40\%/\sqrt{N}$ for binary neutron stars, and upwards of $\sim100\%/\sqrt{N}$ for binary black holes, depending on the galaxy catalog completeness and the binary black hole mass function. Some neutron star-black hole mergers are also expected to have counterparts and, depending on the merger rate, may contribute significantly to the standard siren measurement~\cite{PhysRevLett.121.021303}.

Farther in the future, next generation ground-based networks such as the Cosmic Explorer and the Einstein Telescope will detect standard siren sources to very high redshift~\cite{2017PhRvD..95d4024C,PhysRevD.98.023502,2019PhRvD..99f3510Z}. In addition, the space-based LISA gravitational-wave detector will enable revolutionary standard siren measurements of cosmology~\cite{2002luml.conf..207S,2009PhRvD..80j4009C,2019BAAS...51c..67C}.

After thirty years of development, standard sirens are now poised to become an important new tool in our exploration of the universe. It will be exciting to see what the next thirty years will bring.

\bibliographystyle{plain}
\bibliography{references}

\begin{thebibliography}{10}

\bibitem{ObsScen}
B.~P. {Abbott}, R.~{Abbott}, T.~D. {Abbott}, M.~R. {Abernathy}, F.~{Acernese},
  K.~{Ackley}, and {Adams}.
\newblock {Prospects for observing and localizing gravitational-wave transients
  with Advanced LIGO, Advanced Virgo and KAGRA}.
\newblock {\em Living Reviews in Relativity}, 21(1):3, Apr 2018.

\bibitem{2017Natur.551...85A}
B.~P. {Abbott}, R.~{Abbott}, T.~D. {Abbott}, F.~{Acernese}, K.~{Ackley},
  C.~{Adams}, T.~{Adams}, P.~{Addesso}, R.~X. {Adhikari}, and V.~B. {Adya}.
\newblock {A gravitational-wave standard siren measurement of the Hubble
  constant}.
\newblock {\em \nat}, 551(7678):85--88, Nov 2017.

\bibitem{2017ApJ...848L..13A}
B.~P. {Abbott}, R.~{Abbott}, T.~D. {Abbott}, F.~{Acernese}, K.~{Ackley},
  C.~{Adams}, T.~{Adams}, P.~{Addesso}, R.~X. {Adhikari}, and V.~B. {Adya}.
\newblock {Gravitational Waves and Gamma-Rays from a Binary Neutron Star
  Merger: GW170817 and GRB 170817A}.
\newblock {\em \apjl}, 848(2):L13, Oct 2017.

\bibitem{2017ApJ...848L..12A}
B.~P. {Abbott}, R.~{Abbott}, T.~D. {Abbott}, F.~{Acernese}, K.~{Ackley},
  C.~{Adams}, T.~{Adams}, P.~{Addesso}, R.~X. {Adhikari}, and V.~B. {Adya}.
\newblock {Multi-messenger Observations of a Binary Neutron Star Merger}.
\newblock {\em \apjl}, 848(2):L12, Oct 2017.

\bibitem{PhysRevLett.119.161101}
B.~P. Abbott and et~al.
\newblock Gw170817: Observation of gravitational waves from a binary neutron
  star inspiral.
\newblock {\em Phys. Rev. Lett.}, 119:161101, Oct 2017.

\bibitem{Planck:2015}
P.~A.~R. {Ade}, N.~{Aghanim}, M.~{Arnaud}, M.~{Ashdown}, J.~{Aumont}, and
  {Baccigalupi}.
\newblock {Planck 2015 results. XIII. Cosmological parameters}.
\newblock {\em \aap}, 594:A13, Sep 2016.

\bibitem{Amendola2017}
L.~{Amendola}, I.~{Sawicki}, M.~{Kunz}, and I.~D. {Saltas}.
\newblock {Direct detection of gravitational waves can measure the time
  variation of the Planck mass}.
\newblock {\em ArXiv e-prints}, December 2017.

\bibitem{Belgacem2017}
E.~{Belgacem}, Y.~{Dirian}, S.~{Foffa}, and M.~{Maggiore}.
\newblock {The gravitational-wave luminosity distance in modified gravity
  theories}.
\newblock {\em ArXiv e-prints}, December 2017.

\bibitem{2017PhRvD..95d4024C}
Rong-Gen {Cai} and Tao {Yang}.
\newblock {Estimating cosmological parameters by the simulated data of
  gravitational waves from the Einstein Telescope}.
\newblock {\em \prd}, 95(4):044024, Feb 2017.

\bibitem{2019BAAS...51c..67C}
Robert {Caldwell}, Mustafa {Amin}, Craig {Hogan}, Kelly {Holley-Bockelmann},
  Daniel {Holz}, Philippe {Jetzer}, Ely {Kovitz}, Priya {Natarajan}, David
  {Shoemaker}, and Tristan {Smith}.
\newblock {Astro2020 Science White Paper: Cosmology with a Space-Based
  Gravitational Wave Observatory}.
\newblock In {\em \baas}, volume~51, page~67, May 2019.

\bibitem{chen17}
Hsin-Yu {Chen}, Maya {Fishbach}, and Daniel~E. {Holz}.
\newblock {A two per cent Hubble constant measurement from standard sirens
  within five years}.
\newblock {\em \nat}, 562:545--547, Oct 2018.

\bibitem{Coulter1556}
D.~A. Coulter, R.~J. Foley, C.~D. Kilpatrick, M.~R. Drout, A.~L. Piro, B.~J.
  Shappee, M.~R. Siebert, J.~D. Simon, N.~Ulloa, D.~Kasen, B.~F. Madore,
  A.~Murguia-Berthier, Y.-C. Pan, J.~X. Prochaska, E.~Ramirez-Ruiz, A.~Rest,
  and C.~Rojas-Bravo.
\newblock Swope supernova survey 2017a (sss17a), the optical counterpart to a
  gravitational wave source.
\newblock 358(6370):1556--1558, 2017.

\bibitem{2009PhRvD..80j4009C}
Curt {Cutler} and Daniel~E. {Holz}.
\newblock {Ultrahigh precision cosmology from gravitational waves}.
\newblock {\em \prd}, 80(10):104009, Nov 2009.

\bibitem{2006PhRvD..74f3006D}
Neal {Dalal}, Daniel~E. {Holz}, Scott~A. {Hughes}, and Bhuvnesh {Jain}.
\newblock {Short GRB and binary black hole standard sirens as a probe of dark
  energy}.
\newblock {\em \prd}, 74(6):063006, Sep 2006.

\bibitem{2018MNRAS.479.2374D}
G.~{D{\'a}lya}, G.~{Galg{\'o}czi}, L.~{Dobos}, Z.~{Frei}, I.~S. {Heng},
  R.~{Macas}, C.~{Messenger}, P.~{Raffai}, and R.~S. {de Souza}.
\newblock {GLADE: A galaxy catalogue for multimessenger searches in the
  advanced gravitational-wave detector era}.
\newblock {\em \mnras}, 479(2):2374--2381, Sep 2018.

\bibitem{delpozzo}
W.~{Del Pozzo}.
\newblock {Inference of cosmological parameters from gravitational waves:
  Applications to second generation interferometers}.
\newblock {\em \prd}, 86(4):043011, August 2012.

\bibitem{2017PhRvD..95d3502D}
Walter {Del Pozzo}, Tjonnie G.~F. {Li}, and Chris {Messenger}.
\newblock {Cosmological inference using only gravitational wave observations of
  binary neutron stars}.
\newblock {\em \prd}, 95(4):043502, Feb 2017.

\bibitem{2018PhRvD..98h3523D}
Eleonora {Di Valentino}, Daniel~E. {Holz}, Alessand~ro {Melchiorri}, and
  Fabrizio {Renzi}.
\newblock {Cosmological impact of future constraints on H$_{0}$ from
  gravitational-wave standard sirens}.
\newblock {\em \prd}, 98(8):083523, Oct 2018.

\bibitem{2018arXiv180203404F}
S.~M. {Feeney}, H.~V. {Peiris}, A.~R. {Williamson}, S.~M. {Nissanke}, D.~J.
  {Mortlock}, J.~{Alsing}, and D.~{Scolnic}.
\newblock {Prospects for resolving the Hubble constant tension with standard
  sirens}.
\newblock {\em ArXiv e-prints}, February 2018.

\bibitem{1996PhRvD..53.2878F}
Lee~Samuel {Finn}.
\newblock {Binary inspiral, gravitational radiation, and cosmology}.
\newblock {\em \prd}, 53(6):2878--2894, Mar 1996.

\bibitem{2019ApJ...871L..13F}
M.~{Fishbach}, R.~{Gray}, I.~{Maga{\~n}a Hernandez}, H.~{Qi}, A.~{Sur},
  F.~{Acernese}, L.~{Aiello}, A.~{Allocca}, M.~A. {Aloy}, and A.~{Amato}.
\newblock {A Standard Siren Measurement of the Hubble Constant from GW170817
  without the Electromagnetic Counterpart}.
\newblock {\em \apjl}, 871(1):L13, Jan 2019.

\bibitem{2017ApJ...851L..36G}
C.~{Guidorzi}, R.~{Margutti}, D.~{Brout}, D.~{Scolnic}, W.~{Fong}, K.~D.
  {Alexander}, P.~S. {Cowperthwaite}, J.~{Annis}, E.~{Berger}, and P.~K.
  {Blanchard}.
\newblock {Improved Constraints on H $_{0}$ from a Combined Analysis of
  Gravitational-wave and Electromagnetic Emission from GW170817}.
\newblock {\em \apjl}, 851(2):L36, Dec 2017.

\bibitem{2005ApJ...629...15H}
D.~E. {Holz} and S.~A. {Hughes}.
\newblock {Using Gravitational-Wave Standard Sirens}.
\newblock {\em \apj}, 629:15--22, August 2005.

\bibitem{2018PhT....71l..34H}
Daniel~E. {Holz}, Scott~A. {Hughes}, and Bernard~F. {Schutz}.
\newblock {Measuring cosmic distances with standard sirens}.
\newblock {\em Physics Today}, 71(12):34--40, Dec 2018.

\bibitem{2018arXiv180610596H}
Kenta {Hotokezaka}, Ehud {Nakar}, Ore {Gottlieb}, Samaya {Nissanke}, Kento
  {Masuda}, Gregg {Hallinan}, Kunal~P. {Mooley}, and Adam.~T. {Deller}.
\newblock {A Hubble constant measurement from superluminal motion of the jet in
  GW170817}.
\newblock {\em arXiv e-prints}, page arXiv:1806.10596, Jun 2018.

\bibitem{2019PhRvD..99h3504L}
Macarena {Lagos}, Maya {Fishbach}, Philippe {Landry}, and Daniel~E. {Holz}.
\newblock {Standard sirens with a running Planck mass}.
\newblock {\em \prd}, 99(8):083504, Apr 2019.

\bibitem{Linder2018}
Eric~V. {Linder}.
\newblock {No slip gravity}.
\newblock {\em \jcap}, 2018(3):005, Mar 2018.

\bibitem{macleod}
C.~L. {MacLeod} and C.~J. {Hogan}.
\newblock {Precision of Hubble constant derived using black hole binary
  absolute distances and statistical redshift information}.
\newblock {\em \prd}, 77(4):043512, February 2008.

\bibitem{2012PhRvL.108i1101M}
C.~{Messenger} and J.~{Read}.
\newblock {Measuring a Cosmological Distance-Redshift Relationship Using Only
  Gravitational Wave Observations of Binary Neutron Star Coalescences}.
\newblock {\em \prl}, 108(9):091101, Mar 2012.

\bibitem{PhysRevD.98.023502}
Remya Nair, Sukanta Bose, and Tarun~Deep Saini.
\newblock Measuring the hubble constant: Gravitational wave observations meet
  galaxy clustering.
\newblock {\em Phys. Rev. D}, 98:023502, Jul 2018.

\bibitem{Nishizawa2017}
A.~{Nishizawa}.
\newblock {Generalized framework for testing gravity with gravitational-wave
  propagation. I. Formulation}.
\newblock {\em ArXiv e-prints}, October 2017.

\bibitem{2013arXiv1307.2638N}
S.~{Nissanke}, D.~E. {Holz}, N.~{Dalal}, S.~A. {Hughes}, J.~L. {Sievers}, and
  C.~M. {Hirata}.
\newblock {Determining the Hubble constant from gravitational wave observations
  of merging compact binaries}.
\newblock {\em ArXiv e-prints}, July 2013.

\bibitem{2010ApJ...725..496N}
S.~{Nissanke}, D.~E. {Holz}, S.~A. {Hughes}, N.~{Dalal}, and J.~L. {Sievers}.
\newblock {Exploring Short Gamma-ray Bursts as Gravitational-wave Standard
  Sirens}.
\newblock {\em \apj}, 725:496--514, December 2010.

\bibitem{2018JCAP...07..048P}
Kris {Pardo}, Maya {Fishbach}, Daniel~E. {Holz}, and David~N. {Spergel}.
\newblock {Limits on the number of spacetime dimensions from GW170817}.
\newblock {\em \jcap}, 2018(7):048, Jul 2018.

\bibitem{shoes}
A.~G. {Riess}, L.~M. {Macri}, S.~L. {Hoffmann}, D.~{Scolnic}, S.~{Casertano},
  A.~V. {Filippenko}, B.~E. {Tucker}, M.~J. {Reid}, D.~O. {Jones}, J.~M.
  {Silverman}, R.~{Chornock}, P.~{Challis}, W.~{Yuan}, P.~J. {Brown}, and R.~J.
  {Foley}.
\newblock {A 2.4\% Determination of the Local Value of the Hubble Constant}.
\newblock {\em \apj}, 826:56, July 2016.

\bibitem{schutz}
B.~F. {Schutz}.
\newblock {Determining the Hubble constant from gravitational wave
  observations}.
\newblock {\em \nat}, 323:310, September 1986.

\bibitem{2002luml.conf..207S}
Bernard~F. {Schutz}.
\newblock {Lighthouses of Gravitational Wave Astronomy}.
\newblock In Marat {Gilfanov}, Rashid {Sunyeav}, and Eugene {Churazov},
  editors, {\em Lighthouses of the Universe: The Most Luminous Celestial
  Objects and Their Use for Cosmology}, page 207, Jan 2002.

\bibitem{2011CQGra..28l5023S}
Bernard~F. {Schutz}.
\newblock {Networks of gravitational wave detectors and three figures of
  merit}.
\newblock {\em Classical and Quantum Gravity}, 28(12):125023, Jun 2011.

\bibitem{2014ApJ...795..105S}
Leo~P. {Singer}, Larry~R. {Price}, Ben {Farr}, Alex~L. {Urban}, Chris {Pankow},
  Salvatore {Vitale}, John {Veitch}, Will~M. {Farr}, Chad {Hanna}, and Kipp
  {Cannon}.
\newblock {The First Two Years of Electromagnetic Follow-up with Advanced LIGO
  and Virgo}.
\newblock {\em \apj}, 795(2):105, Nov 2014.

\bibitem{2017ApJ...848L..16S}
M.~{Soares-Santos}, D.~E. {Holz}, J.~{Annis}, R.~{Chornock}, K.~{Herner},
  E.~{Berger}, D.~{Brout}, H.~Y. {Chen}, R.~{Kessler}, and M.~{Sako}.
\newblock {The Electromagnetic Counterpart of the Binary Neutron Star Merger
  LIGO/Virgo GW170817. I. Discovery of the Optical Counterpart Using the Dark
  Energy Camera}.
\newblock {\em \apjl}, 848(2):L16, Oct 2017.

\bibitem{2019ApJ...876L...7S}
M.~{Soares-Santos}, A.~{Palmese}, W.~{Hartley}, J.~{Annis},
  J.~{Garcia-Bellido}, O.~{Lahav}, Z.~{Doctor}, M.~{Fishbach}, D.~E. {Holz},
  and H.~{Lin}.
\newblock {First Measurement of the Hubble Constant from a Dark Standard Siren
  using the Dark Energy Survey Galaxies and the LIGO/Virgo Binary─Black-hole
  Merger GW170814}.
\newblock {\em \apjl}, 876(1):L7, May 2019.

\bibitem{2012PhRvD..85b3535T}
Stephen~R. {Taylor}, Jonathan~R. {Gair}, and Ilya {Mandel}.
\newblock {Cosmology using advanced gravitational-wave detectors alone}.
\newblock {\em \prd}, 85(2):023535, Jan 2012.

\bibitem{PhysRevLett.121.021303}
Salvatore Vitale and Hsin-Yu Chen.
\newblock Measuring the hubble constant with neutron star black hole mergers.
\newblock {\em Phys. Rev. Lett.}, 121:021303, Jul 2018.

\bibitem{2019PhRvD..99f3510Z}
Xuan-Neng {Zhang}, Ling-Feng {Wang}, Jing-Fei {Zhang}, and Xin {Zhang}.
\newblock {Improving cosmological parameter estimation with the future
  gravitational-wave standard siren observation from the Einstein Telescope}.
\newblock {\em \prd}, 99(6):063510, Mar 2019.

\end{thebibliography}

\vfill\eject

\section*{\centerline {Hartlefest}}
\addtocontents{toc}{\protect\medskip}
\addtocontents{toc}{\bf Conference Reports:}
\addcontentsline{toc}{subsubsection}{
\it Hartlefest, by Gary Horowitz}
\parskip=3pt
\begin{center}
Gary Horowitz, University of California Santa Barbara 
\htmladdnormallink{horowitz-at-ucsb.edu}
{mailto:horowitz@ucsb.edu}
\end{center}

On June 7, 2019, approximately 70 physicists gathered at the KITP in Santa Barbara for a one-day celebration of Jim Hartle's 80th birthday. There were seven talks: Kip Thorne and Gary Gibbons reviewed Jim's many contributions to physics and astrophysics going back to the 1960s; Neil Turok and Thomas Hertog discussed various aspects of the Hartle-Hawking no-boundary wave function; Sean Carroll and Mark Srednicki discussed aspects of probability in quantum cosmology and complexity; and a philosopher, Simon Saunders, discussed the many worlds view of quantum mechanics. Many distinguished relativists attended this event, including Bob Wald, John Friedman, Jim Bardeen, and many others (including three of Jim's former students: Paul Anderson, David Craig and Peter Morse). For a complete list of participants and links to the talks and pictures, see the conference website: 
\htmladdnormallink 
{\protect {\tt {http://web.physics.ucsb.edu/~HartleFest/}}}
{http://web.physics.ucsb.edu/~HartleFest/} 

After dinner, there were reminiscences and remarks about how kind and helpful Jim has been to generations of students and postdocs by a number of people including Wald, Bardeen, Isaacson, Zurek, and a couple of Jim's former students. It was a special event honoring a very special person. 

\end{document}